\documentclass[nohyperref]{article}

\usepackage{microtype}
\usepackage{graphicx}
\usepackage{subfigure}
\usepackage{booktabs} % for professional tables

\usepackage{hyperref}
\usepackage[accepted]{icml2022}

\usepackage{amsmath}
\usepackage{amssymb}
\usepackage{mathtools}
\usepackage{amsthm}
\usepackage{multirow}

\usepackage[capitalize,noabbrev]{cleveref}

\theoremstyle{plain}

\theoremstyle{definition}

\theoremstyle{remark}

\usepackage[textsize=tiny]{todonotes}

\icmltitlerunning{Generating 3D Molecules for Target Protein Binding}

\begin{document}

\twocolumn[
% \icmltitle{Placing Atoms to 3D Binding Sites for Drug Design}
\icmltitle{Generating 3D Molecules for Target Protein Binding}




\begin{icmlauthorlist}
\icmlauthor{Meng Liu}{tamu}
\icmlauthor{Youzhi Luo}{tamu}
\icmlauthor{Kanji Uchino}{fujitsu_us}
\icmlauthor{Koji Maruhashi}{fujitsu_jp}
\icmlauthor{Shuiwang Ji}{tamu}
\end{icmlauthorlist}

\icmlaffiliation{tamu}{Department of Computer Science \& Engineering, Texas A\&M University, TX, USA}
\icmlaffiliation{fujitsu_us}{Fujitsu Research of America, INC., CA, USA}
\icmlaffiliation{fujitsu_jp}{Fujitsu Research, Fujitsu Limited, Kanagawa, Japan}

% \icmlcorrespondingauthor{Meng Liu}{mengliu@tamu.edu}
% \icmlcorrespondingauthor{Kanji Uchino}{kanji@fujitsu.com}
\icmlcorrespondingauthor{Shuiwang Ji}{sji@tamu.edu}

\icmlkeywords{Machine Learning, ICML}

\vskip 0.3in
]


\printAffiliationsAndNotice{}  % leave blank if no need to mention equal contribution

% the abs for ICML abs deadline
% Background --> overview --> key steps
% \begin{abstract}
% A fundamental problem in drug discovery is to design molecules that bind to specific proteins.
% To tackle this problem using machine learning methods, here we propose a novel and effective framework to generate 3D molecules that bind to given proteins by placing atoms of specific types and locations to the given binding site one by one. In particular, at each step, we first employ a 3D graph neural network to obtain geometry-aware and chemically informative representations from the intermediate contextual information. Such context includes the given binding site and atoms placed in the previous steps. Second, to preserve the desirable equivariance property, we select a local reference atom according to the designed auxiliary classifiers and then construct a local spherical coordinate system. Finally, to place a new atom, we generate its atom type and relative location \emph{w.r.t.} the constructed local coordinate system via a flow model. We also consider generating the variables of interest sequentially to capture the underlying dependencies among them. Experiments demonstrate that our proposed approach is effective to generate 3D molecules with binding ability to target protein binding sites.
% \end{abstract}

\begin{abstract}
A fundamental problem in drug discovery is to design molecules that bind to specific proteins.
To tackle this problem using machine learning methods, here we propose a novel and effective framework, known as GraphBP, to generate 3D molecules that \underline{b}ind to given \underline{p}roteins by placing atoms of specific types and locations to the given binding site one by one. In particular, at each step, we first employ a 3D graph neural network to obtain geometry-aware and chemically informative representations from the intermediate contextual information. Such context includes the given binding site and atoms placed in the previous steps. Second, to preserve the desirable equivariance property, we select a local reference atom according to the designed auxiliary classifiers and then construct a local spherical coordinate system. Finally, to place a new atom, we generate its atom type and relative location \emph{w.r.t.} the constructed local coordinate system via a flow model. We also consider generating the variables of interest sequentially to capture the underlying dependencies among them. Experiments demonstrate that our GraphBP is effective to generate 3D molecules with binding ability to target protein binding sites. Our implementation is available at \url{https://github.com/divelab/GraphBP}.
\end{abstract}

\section{Introduction}
\label{sec:intro}

% Intro + machine learning methods are promising but under-explored for this problem
Designing molecules that can bind to a specific target protein (\emph{a.k.a.} structure-based drug design) is a fundamental and challenging problem in drug discovery~\citep{anderson2003process}. It is highly promising to develop machine learning methods for this problem since there are recently available large-scale datasets of protein-ligand complex structures, such as PDBbind~\citep{liu2017forging} and CrossDocked2020~\citep{francoeur2020three}. In addition, machine learning approaches have been shown to be effective for learning from richly structured data in biochemistry. The most representative example is AlphaFold~\citep{jumper2021highly}, which achieves remarkable accuracy on the problem of 3D protein structure prediction from amino acid sequence, a long-standing challenge for decades.

% Key challenges and considerations: (1) Conditional on receptors: in terms of both geometric and chemical interactions; (2) Enormous chemical space and continuous 3D space; (3) Rotational and translational equivariance
However, machine learning approaches have rarely been explored to generate molecules that bind to specific protein binding sites. We summarize the main challenges or considerations in three folds. (i) \textbf{Complicated conditional information.} When generating molecules that are capable of binding to a specific target protein, both the 3D geometric structure and the chemical features of the binding site are important considerations. It is crucial to consider how to capture such informative context effectively. (ii) \textbf{Enormous chemical space and continuous 3D space.} The chemical space of all possible molecules is enormous (estimated to be larger than $10^{60}$), while the number of molecules that have binding ability to a specific target is extremely small. In addition, the 3D space around the binding site is continuous by nature. In other words, it is desirable that our generative model is capable of generating molecules in any continuous positions without discretizing the space. (iii) \textbf{Equivariance property.} Intuitively, if we rotate or translate the binding site, the generated molecules are expected to be rotated or translated the same way. That is, molecules generated by our machine learning approach should be equivariant to any rigid transformation of the binding site.

% This work: how this method considers the above properties. 

Here, we present GraphBP, a novel and effective generative framework for structure-based drug design, that takes the described challenges into consideration. Particularly, we generate 3D molecules by placing atoms to the specific 3D binding site one by one. At each step, a 3D graph neural network is firstly employed to extract the intermediate contextual information by considering both 3D geometric structures and chemical interactions. Afterwards, we construct a local coordinate system based on a local reference atom selected by the designed auxiliary classifiers. Generating a new atom in this local coordinate system can ensure the equivariance property. Finally, to place a new atom, we generate its atom type and relative continuous position \emph{w.r.t.} the constructed local coordinate system with a flow model. Moreover, the variables of interest are generated sequentially, aiming to capture the underlying dependencies.

% Our pros over prior works
To our knowledge, in structure-based drug design, our GraphBP is the first machine learning method that satisfies all of the following three characteristics; that is, it can perceive 3D geometric structures and chemical interactions of protein-ligand complexes, place atoms in any continuous positions, and preserve the desirable equivariance property. More discussions with prior works~\citep{ragoza2021generating,luo2021sbdd} are included in Section~\ref{sec:related_works}. Experiments show that our approach outperforms baselines significantly in generating 3D molecules that have binding affinity to target 3D protein binding sites.

% Experiments

\section{Preliminaries and Related Work}
\label{sec:related_works}

% 1D/2D molecule generation methods; unable to perceive 3D structures
% longer version
% \textbf{1D/2D molecule generation.} Molecules can be represented as 1D SMILES strings~\citep{weininger1988smiles} or 2D molecular graphs. Several works propose to generate SMILES strings~\citep{gomez2018automatic,kusner2017grammar,dai2018syntax} with sequence generative methods. Alternatively, many works generate 2D graphs by leveraging advanced deep generative models, such as VAEs~\citep{kingma2013auto,simonovsky2018graphvae,liu2018constrained,ma2018constrained,jin2018junction}, GANs~\citep{goodfellow2014generative,de2018molgan,you2018graph}, flow models~\citep{dinh2014nice,madhawa2019graphnvp,liu2019graph,shi2019graphaf,zang2020moflow,lippe2020categorical,luo2021graphdf}, RNNs~\citep{li2018learning,you2018graphrnn,popova2019molecularrnn}, and energy-based models~\citep{lecun2006tutorial,liu2021graphebm}. These methods generate 1D or 2D molecules without perceiving 3D spatial information. Thus, they cannot be directly applied to generate 3D molecules for target protein binding.

\textbf{1D/2D molecule generation.} Molecules can be represented as 1D SMILES strings~\citep{weininger1988smiles} or 2D molecular graphs. Several works propose to generate SMILES strings~\citep{gomez2018automatic,kusner2017grammar,dai2018syntax} with sequence methods. Alternatively, many works generate 2D graphs by leveraging advanced deep generative models. They either generate the node type matrix and adjacency matrix directly~\citep{simonovsky2018graphvae,de2018molgan,zang2020moflow,liu2021graphebm}, or generate nodes, edges, or motifs by adding them one by one~\citep{li2018learning,you2018graphrnn,jin2018junction,shi2019graphaf,luo2021graphdf}. These methods generate 1D or 2D molecules without perceiving 3D spatial information. Thus, they cannot be directly applied to generate 3D molecules for target protein binding.

% 3D molecule generation (from scratch or 2D input); unable to model the interaction with binding sites
\textbf{3D molecule generation.} Recently, many works propose to generate 3D molecular geometries from given 2D graphs~\cite{mansimov2019molecular,simm2020generative,gogineni2020torsionnet,xu2021end,shi2021confgf,ganea2021geomol,luo2021predicting}, from a given bag of atoms~\citep{simm2020reinforcement}, or from scratch~\citep{gebauer2019symmetry,hoffmann2019generating,nesterov20203dmolnet,satorras2021n,luo2022gspherenet}. In structure-based drug design, however, the prior knowledge of 2D graphs or the bag of atoms are unknown. In addition, these methods usually consider small organic molecules~\citep{luo2021sbdd}, thus remaining to be insufficient to generate 3D drug-like molecules interacting with given binding sites. For a comprehensive review of molecule generation, we recommend referring to the recent survey~\citep{du2022molgensurvey}.

% Structure-based drug design: String-based, LiGAN and LuoST. Pros over previous work
\textbf{Structure-based drug design.} Generating 3D molecules that bind to specific binding sites with machine learning approaches is challenging and under-explored. LiGAN~\citep{ragoza2021generating} converts protein-ligand complexes to 3D atomic density grids, \emph{i.e.}, 3D images. Then it treats structure-based drug design as a 3D image generation task, thus enabling the usage of GANs~\citep{goodfellow2014generative} and VAEs~\citep{kingma2013auto}. After generating density grids, it performs an atom fitting algorithm to obtain 3D molecular geometries. As a preliminary work, it fails to preserve the desirable equivariance property since performing 3D CNNs~\citep{ji20123d} on an atomic density grid is not equivariant. Also, it has to discretize the continuous 3D space to construct grids. Another recent work~\citep{luo2021sbdd} tackles this problem by modeling the distribution of atom occurrence in the 3D space around the binding site, and then employing a sampling algorithm to place atoms according to the learned distribution. During sampling, it also discretizes the 3D space onto meshgrids and evaluates the probability densities of atom’s occurrences on the grids. In contrast, our method can place the atoms in any continuous positions, thereby enabling more flexible atom placement.

% Preliminary: flow models --> autoregressive flow models
\textbf{Autoregressive flow models.} A flow model~\citep{dinh2014nice,rezende2015variational,weng2018flow} defines a parameterized invertible  transformation function $f_\theta:\boldsymbol{z}\in \mathbb{R}^D \rightarrow \boldsymbol{x}\in \mathbb{R}^D$ from latent variable $\boldsymbol{z} \sim p_{Z}$ to data variable $\boldsymbol{x}$, where $p_{Z}$ is a known prior distribution. The log-likelihood of a data point $\boldsymbol{x}$ can be computed by
\begin{equation}
\label{eq:flow_basic}
    \log p_{X}(\boldsymbol{x}) = \log p_{Z}\left(f_\theta^{-1}(\boldsymbol{x})\right) + \log \left|\text{det}\frac{\partial f_\theta^{-1}(\boldsymbol{x})}{\partial \boldsymbol{x}}\right|.
\end{equation}
Thus, $f_\theta$ is required to be invertible and its Jacobian determinant should be computed easily. An autoregressive flow model~\citep{papamakarios2017masked} is a specific flow method where the transformation function is formulated as an autoregressive model; that is, each dimension of $\boldsymbol{x}$ is conditioned on the previous dimensions. Formally, it is usually defined as an affine transformation as
\begin{equation}
    \boldsymbol{x}_i = \sigma_i(\boldsymbol{x}_{1:i-1})\odot \boldsymbol{z}_i + \mu_i(\boldsymbol{x}_{1:i-1}), \quad i=1,\cdots,D,
\end{equation}
where the scale factor $\sigma_i(\cdot) \in \mathbb{R}$ and the translation factor $\mu_i(\cdot) \in \mathbb{R}$ are functions of $\boldsymbol{x}_{1:i-1}$. $\odot$ denotes the element-wise multiplication. This transformation function is easy to inverse as $\boldsymbol{z}_i=\frac{\boldsymbol{x}_i - \mu_i}{\sigma_i}$. In addition, the determinant of the Jacobian matrix can be computed linearly since it is a triangular matrix. To be specific, $\text{det}\frac{\partial f_\theta^{-1}(\boldsymbol{x})}{\partial \boldsymbol{x}}=\prod_{i=1}^D\frac{1}{\sigma_i}$.

\section{Method}

% Notation and problem
\textbf{Notations and problem.} We represent the 3D geometry of a molecule (\emph{i.e.}, a ligand) as $\mathcal{M}=\{(\boldsymbol{a}_i,\boldsymbol{r}_i)\}_{i=1}^n$ and the corresponding binding site of a protein (\emph{i.e.}, a receptor) as $\mathcal{P}=\{(\boldsymbol{b}_j,\boldsymbol{s}_j)\}_{j=1}^m$. $n$ and $m$ denote the numbers of atoms in the molecule and in the binding site, respectively. $\boldsymbol{a}_i \in \{0,1\}^p$ is the one-hot vector indicating the atom type of the $i$-th atom in the molecule, and $\boldsymbol{r}_i \in \mathbb{R}^3$ is its 3D Cartesian coordinate. Similarly, the atom type and the coordinate of the $j$-th atom in the binding site are denoted as one-hot vector $\boldsymbol{b}_j \in \{0,1\}^q$ and $\boldsymbol{s}_j \in \mathbb{R}^3$. $p$ and $q$ represent the total numbers of atom types in molecules and in binding sites, respectively, and they can be obtained from the statistics of the training set. We consider the problem of generating 3D molecules in the given binding site. Thus, our goal is to learn a generative model to capture the conditional distribution $p(\mathcal{M}|\mathcal{P})$ of observed protein-ligand pairs.

\begin{figure*}[t]
	\centering
	\includegraphics[width=\textwidth]{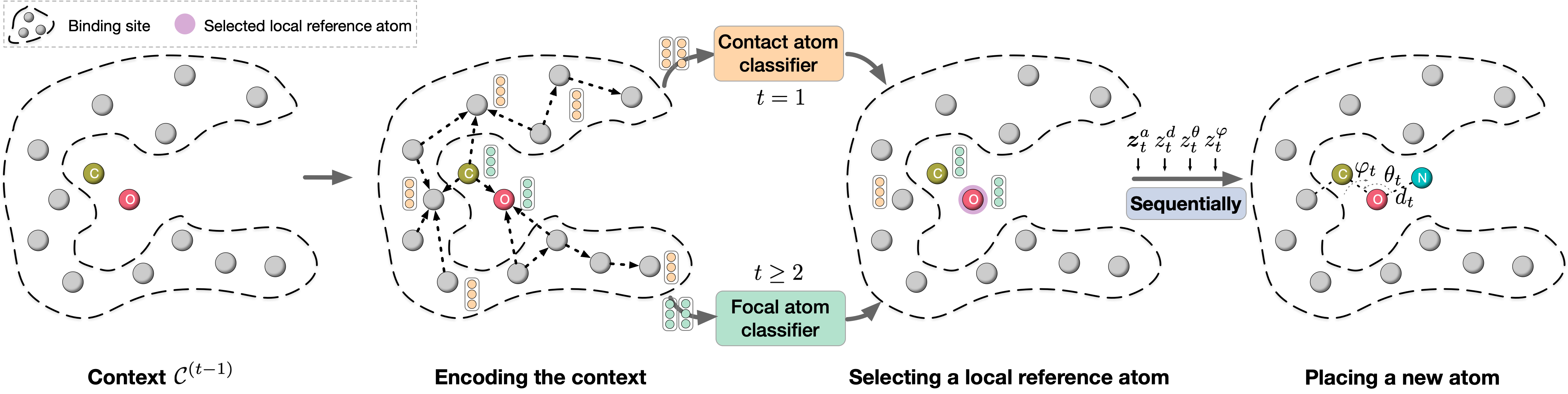}
	\caption{An illustration of one generation step of GraphBP. Details are described in Section~\ref{sec:gen}.}
	\label{fig:overview_gen}
\end{figure*}

\subsection{Generation}
\label{sec:gen}

% Overview of the sequential generation process (with an illustration figure)
\textbf{Overview.} In GraphBP, we formulate the generation of 3D molecules in the given binding site as a sequential generation process; that is, we place atoms to the given 3D binding site one by one. At the $t$-th step, we generate the $t$-th atom, including its atom type $\boldsymbol{a}_t$ and coordinate $\boldsymbol{r}_t$, based on the intermediate contextual information $\mathcal{C}^{(t-1)}$. Note that the context $\mathcal{C}^{(t-1)}$ contains not only the binding site but also the atoms placed in the previous $t-1$ steps, \emph{i.e.}, $\mathcal{C}^{(t-1)} = \mathcal{P} \cup \{(\boldsymbol{a}_i,\boldsymbol{r}_i)\}_{i=1}^{t-1}$ when $t\geq 2$. At the first step ($t=1$), the context is the binding site itself, \emph{i.e.}, $\mathcal{C}^{(0)}=\mathcal{P}$.

Within each step, we firstly generate the atom type based on the context. Afterwards, its coordinate is generated by considering both the context and the generated atom type information. Therefore, each step $t\ (t=1,2,\cdots,n)$ of our generation process can be formulated as
\begin{equation}
\label{eq:sequential_gen}
    \begin{aligned}
    &\boldsymbol{a}_t = g^a\left(\mathcal{C}^{(t-1)}; \boldsymbol{z}_t^a\right), \\
    &\boldsymbol{r}_t = g^r\left(\mathcal{C}^{(t-1)}, \boldsymbol{a}_t; \boldsymbol{z}_t^r\right), \\
    &\mathcal{C}^{(t)} \leftarrow \mathcal{C}^{(t-1)} \cup \left\{(\boldsymbol{a}_t,\boldsymbol{r}_t)\right\}.
    \end{aligned}
\end{equation}
Generators $g^a(\cdot)$ and $g^r(\cdot)$ are autoregressive functions. $\boldsymbol{z}_t^a$ and $\boldsymbol{z}_t^r$ denote the latent variables used in the flow model at step $t$, which will be introduced in details later.

In the following, we describe the details of one generation step, \emph{i.e.}, how the autoregressive functions $g^a(\cdot)$ and $g^r(\cdot)$ are parameterized. In addition, we also explain how the key challenges summarized in Section~\ref{sec:intro} are considered in GraphBP. Particularly, there are mainly three parts in one generation step, namely \textbf{encoding the context}, \textbf{selecting a local reference atom}, and \textbf{placing a new atom}, as illustrated in Figure~\ref{fig:overview_gen}. The details are elucidated as follows.

% General: why 3D GNN
% \textbf{Step 1: Encoding the context.} 
\subsubsection{Encoding the Context}
\label{sec:step1}

As introduced in Section~\ref{sec:intro}, both geometric shape and chemical interactions are vital to protein-ligand binding affinity. Hence, it is important to capture such information by the context encoder. We firstly construct a graph for the context $\mathcal{C}^{(t-1)}$ by connecting atoms with considering certain cutoff distance. Let $\mathcal{G}^{(t-1)}$ denote the obtained context graph. Afterwards, we employ a 3D graph neural network (3D GNN) to encode $\mathcal{G}^{(t-1)}$. Formally,
\begin{equation}
\label{eq:3DGNN}
    \{\boldsymbol{h}_1^{(t)}, \cdots, \boldsymbol{h}_{m+t-1}^{(t)}\} = \text{3DGNN}\left(\mathcal{G}^{(t-1)}\right),
\end{equation}
where $\boldsymbol{h}_k^{(t)}$ represents the encoded representation of atom $k$ in the context $\mathcal{C}^{(t-1)}$. Note that there are totally $m+t-1$ atoms in the context, including $m$ atoms from the binding site and $t-1$ atoms placed in the previous $t-1$ steps.

% To achieve this, we can naturally adapt the recently developed 3D graph neural networks (3D GNNs)~\cite{schlichtkrull2018modeling,klicpera2019directional,liu2021spherical,hu2021forcenet,klicpera_gemnet_2021}, which can consider both 3D geometric information and chemical interactions in feature aggregation, to extract informative features from the context.

% Details
The first layer of our 3DGNN is an embedding layer for encoding atom types. Note that we use different learnable embeddings for atoms in the binding site and atoms in the ligand, thereby differentiating ligand atoms from protein atoms. For example, a carbon atom in the ligand and a carbon atom in the protein have different initial representations. Let $\{\boldsymbol{h}_1^{(t,0)}, \cdots, \boldsymbol{h}_{m+t-1}^{(t,0)}\}$ be the resulting initial representations. Then, we have $L$ feature aggregation layers in our 3DGNN. The aggregation for each atom $k$ at the $\ell$-th layer ($1 \leq \ell \leq L$) can be formulated as
\begin{equation}
\label{eq:schnet_variant}
    \boldsymbol{h}_k^{(t,\ell)} = \boldsymbol{h}_k^{(t,\ell-1)} + \sum_{u\in \mathcal{N}(k)} \boldsymbol{h}_u^{(t,\ell-1)} \odot \text{MLP}^\ell\left(\boldsymbol{e}_{\text{RBF}}\left(d_{uk}\right)\right),
\end{equation}
where $\mathcal{N}(k)$ denotes neighbors of atom $k$ in $\mathcal{G}^{(t-1)}$, $\text{MLP}^\ell(\cdot)$ is a multi-layer perceptron, and $\odot$ represents the element-wise multiplication. $\boldsymbol{e}_{\text{RBF}}\left(d_{uk}\right)$ is the high-dimensional embedding of the distance $d_{uk}$ using radial basis functions (RBF), such as Gaussian functions~\citep{schlichtkrull2018modeling} and spherical Bessel functions~\citep{klicpera2019directional}. Note that the representations $\{\boldsymbol{h}_1^{(t)}, \cdots, \boldsymbol{h}_{m+t-1}^{(t)}\}$ obtained by these $L$ feature aggregation layers are invariant to the rotation and translation of the context $\mathcal{C}^{(t-1)}$, since the distance $d_{uk}$ used in Eq.~(\ref{eq:schnet_variant}) is rotationally and translationally invariant. Our aggregation layer shown in Eq.~(\ref{eq:schnet_variant}) is a variant of SchNet~\citep{schlichtkrull2018modeling}. We can further employ more advanced but more memory-consuming 3D GNNs~\citep{liu2021DIG}, such as DimeNet~\citep{klicpera2019directional} and SphereNet~\citep{liu2021spherical}, to encode the context information. In this work, we do not use them as our encoder because of insufficient memory budget, given that the context graph could have hundreds of atoms.

% Why local reference: rotation/translation equivariant
% \textbf{Step 2: Selecting a local reference atom.}

\subsubsection{Selecting a Local Reference Atom}

As described in Section~\ref{sec:intro}, the location of a generated molecule should be equivariant to any rigid transformation of the binding site. In other words, if we rotate or translate the binding site, the generated molecule should be rotated or translated correspondingly. In our sequential generation case, as formulated in Eq.~(\ref{eq:sequential_gen}), it is desired that the generated coordinate of the $t$-th atom is equivariant to any rigid transformation of the context $\mathcal{C}^{(t-1)}$, while the generated atom type keeps invariant. Formally,
\begin{equation}
\label{eq:equivariance-invariance}
    \begin{aligned}
    &g^a\left(\mathcal{C}^{(t-1)}; \boldsymbol{z}_t^a\right) = g^a\left(\text{RT}\left(\mathcal{C}^{(t-1)}\right); \boldsymbol{z}_t^a\right), \\
    &\text{RT}\left(g^r\left(\mathcal{C}^{(t-1)}, \boldsymbol{a}_t; \boldsymbol{z}_t^r\right)\right) = g^r\left(\text{RT}\left(\mathcal{C}^{(t-1)}\right), \boldsymbol{a}_t; \boldsymbol{z}_t^r\right),
    \end{aligned}
\end{equation}
where $\text{RT}(\cdot)$ represents any rigid transformation, including rotation, translation, and any composition of them.

As described in Section~\ref{sec:step1}, our atom representations obtained from context encoding are invariant to any rigid transformation of the context. Thus, it is straightforward to generate invariant atom type by using these representations. Nevertheless, it is non-trivial to generate coordinate that are equivariant to any rigid transformation of the context. To achieve this desirable equivariance, inspired by G-SchNet~\citep{gebauer2019symmetry}, MolGym~\citep{simm2020reinforcement}, and G-SphereNet~\citep{luo2022gspherenet}, we choose to construct a local spherical coordinate system (SCS) and generate the invariant 3-tuple $(d_t, \theta_t, \varphi_t)$ \emph{w.r.t.} the constructed local SCS.

% How: $t=1$ and $t>1$
To construct such local SCS, we consider selecting a local reference atom from the context by using two auxiliary atom-wise classifiers; they are \textit{contact atom classifier} (for $t=1$) and \textit{focal atom classifier} (for $t\geq2$). (i) At the first step ($t=1$), the known context information is the binding site. The \textit{contact atom classifier} takes the context-encoded representation of each atom in the binding site as input, and determines if the corresponding atom can serve as a local reference atom (\emph{i.e.}, yes or no). The atom selected based on the \textit{contact atom classifier} will be used as the local reference atom for generating the first atom in the ligand. This selected atom is termed as contact atom because it acts like a ``bridge'' in contact with the ligand. (ii) For $t\geq2$, we select a local reference atom from the ligand atoms generated in the previous $t-1$ steps, considering that the new atom is expected to be placed in the local region of the selected reference atom. To be specific, we apply \textit{focal atom classifier} to the context-encoded representations of all existing atoms in the ligand, which are generated in the previous $t-1$ steps, and classify them into two categories: focal atom and non-focal atom. Then, the selected focal atom will be used as the local reference atom to generate the new atom. Overall, for $t=1$, a local reference atom is selected from the binding site using the \textit{contact atom classifier}. For $t\geq2$, a local reference atom is selected from the existing ligand atoms according to the \textit{focal atom classifier}. We describe how to train these two auxiliary classifiers in Section~\ref{sec:training}.

In general, we need three points in the 3D space to define an SCS. Assuming that the selected local reference atom is the $f$-th atom in the context $\mathcal{C}^{(t-1)}$, we can further find two atoms in the context that are closest and second closest to $f$. These two atoms are denoted as the $c$-th and the $e$-th atom in the context $\mathcal{C}^{(t-1)}$, and they could be in the ligand or in the binding site. With these three atoms $(f, c, e)$, we can construct a local SCS. Further, we can generate the invariant $(d_t, \theta_t, \varphi_t)$ \emph{w.r.t.} this local SCS. Specifically, $d_t$ is distance between the new atom and atom $f$, \emph{i.e.}, $d_t=||\boldsymbol{r}_t-\boldsymbol{r}_f||_2$, $\theta_t \in [0,\pi]$ is the angle between line $(\boldsymbol{r}_f, \boldsymbol{r}_t)$ and line $(\boldsymbol{r}_f, \boldsymbol{r}_c)$, and $\varphi_t \in [-\pi, \pi]$ is the torsion angle formed by plane $(\boldsymbol{r}_f, \boldsymbol{r}_c, \boldsymbol{r}_t)$ and plane $(\boldsymbol{r}_f, \boldsymbol{r}_c, \boldsymbol{r}_e)$. Afterwards, we can compute $\boldsymbol{r}_t$ based on the generated $(d_t, \theta_t, \varphi_t)$ and the known $(\boldsymbol{r}_f, \boldsymbol{r}_c, \boldsymbol{r}_e)$. Note that the constructed local SCS is associated with the context, thus being equivariant to any rigid transformation of the context. In other words, $(\boldsymbol{r}_f, \boldsymbol{r}_c, \boldsymbol{r}_e)$ is equivariant to any rigid transformation of the context. Therefore, the computed $\boldsymbol{r}_t$ also keeps equivariant as long as the generated $(d_t, \theta_t, \varphi_t)$ is invariant to any rigid transformation of the context $\mathcal{C}^{(t-1)}$. In addition, we can achieve flexible atom placement since the generated $(d_t, \theta_t, \varphi_t)$ are continuous values, while previous works~\citep{ragoza2021generating,luo2021sbdd} have to discretize the continuous space during atom placement.

%\textbf{Step 3: Placing a new atom.}

\subsubsection{Placing a New Atom}
\label{sec:step3}

The remaining part in generation is to place a new atom by generating $(d_t, \theta_t, \varphi_t)$ as well as $\boldsymbol{a}_t$. As analyzed above, $\boldsymbol{a}_t$, $d_t$, $\theta_t$, and $\varphi_t$ should be invariant to any rigid transformation of the context $\mathcal{C}^{(t-1)}$. Hence, it is natural to generate them with context-encoded representations of atoms $(f, c, e)$, \emph{i.e.}, $(\boldsymbol{h}_{f}^{(t)}, \boldsymbol{h}_{c}^{(t)}, \boldsymbol{h}_{e}^{(t)})$, which are invariant to the rotation and translation of the context $\mathcal{C}^{(t-1)}$. In addition, intuitively, $\boldsymbol{a}_t$, $d_t$, $\theta_t$, and $\varphi_t$ are not independent to each other. For example, a carbon atom and an oxygen atom have different distributions over the distance to their local reference atoms. Further, atoms with the same atom type but different distances \emph{w.r.t.} their local reference atoms could have different distributions over angles. Thus, we propose to generate $\boldsymbol{a}_t$, $d_t$, $\theta_t$, and $\varphi_t$ sequentially in each generation step to capture their underlying dependencies. To be specific, we generate these variables using the order $\boldsymbol{a}_t \rightarrow d_t \rightarrow \theta_t \rightarrow \varphi_t$, and the generation of each variable is dependent on the previous variables. For instance, to generate $\varphi_t$, in addition to $\mathcal{C}^{(t-1)}$, we incorporate the information of $\boldsymbol{a}_t$, $d_t$, and $\theta_t$. Mathematically, $p\left(\boldsymbol{a}_t, d_t, \theta_t, \varphi_t|\mathcal{C}^{(t-1)}\right)=p\left(\boldsymbol{a}_t|\mathcal{C}^{(t-1)}\right)p\left(d_t|\mathcal{C}^{(t-1)}\right)p\left(\theta_t|\mathcal{C}^{(t-1)}\right)p\left(\varphi_t|\mathcal{C}^{(t-1)}\right)$ does not hold if $\boldsymbol{a}_t$, $d_t$, $\theta_t$, and $\varphi_t$ are not independent. In contrast, the following equation always holds according to the multiplication rule of probability, no matter if the variables are independent or not.
% \begin{equation}
% \begin{aligned}
%     &p\left(\boldsymbol{a}_t, d_t, \theta_t, \varphi_t|\mathcal{C}^{(t-1)}\right) = p\left(\varphi_t|\mathcal{C}^{(t-1)}, \boldsymbol{a}_t, d_t, \theta_t\right) \\
%     &p\left(\theta_t|\mathcal{C}^{(t-1)},\boldsymbol{a}_t, d_t\right)p\left(d_t|\mathcal{C}^{(t-1)},\boldsymbol{a}_t\right)p\left(\boldsymbol{a}_t|\mathcal{C}^{(t-1)}\right).
% \end{aligned}
% \end{equation}
\begin{equation}
\begin{aligned}
    p\left(\boldsymbol{a}_t, d_t, \theta_t, \varphi_t|\mathcal{C}^{(t-1)}\right) = p\left(\boldsymbol{a}_t|\mathcal{C}^{(t-1)}\right)p\left(d_t|\mathcal{C}^{(t-1)},\boldsymbol{a}_t\right)& \\
    p\left(\theta_t|\mathcal{C}^{(t-1)},\boldsymbol{a}_t, d_t\right)p\left(\varphi_t|\mathcal{C}^{(t-1)}, \boldsymbol{a}_t, d_t, \theta_t\right).&
\end{aligned}
\end{equation}
This demonstrates that our generation strategy is also technically sound. We conduct ablation study in Section~\ref{sec:exp} to demonstrate the effectiveness of this sequential generation strategy. Therefore, our one-step generation, as shown in Eq.~(\ref{eq:sequential_gen}), can be reformulated as
\begin{equation}
\label{eq:sequential_gen_reformulated}
    \begin{aligned}
    &\boldsymbol{a}_t = g^a\left(\mathcal{C}^{(t-1)}; \boldsymbol{z}_t^a\right), \\
    &d_t = g^d\left(\mathcal{C}^{(t-1)}, \boldsymbol{a}_t; z_t^d\right), \\
    &\theta_t = g^\theta\left(\mathcal{C}^{(t-1)}, \boldsymbol{a}_t, d_t; z_t^\theta\right), \\
    &\varphi_t = g^\varphi\left(\mathcal{C}^{(t-1)}, \boldsymbol{a}_t, d_t, \theta_t; z_t^\varphi\right),
    \end{aligned}
\end{equation}
where $\boldsymbol{z}_t^a \in \mathbb{R}^p$, $z_t^d \in \mathbb{R}$, $z_t^\theta \in \mathbb{R}$, and $z_t^\varphi \in \mathbb{R}$ are all latent variables used in the flow model. During generation, we sample latent variables from known prior Gaussian distributions, and then map them to variables of interest (\emph{i.e.}, $\boldsymbol{a}_t$, $d_t$, $\theta_t$, $\varphi_t$). For training, we map observed variables to latent variables, and maximize their likelihood. Since the atom type vector is discrete, which cannot fit into a flow model, we convert it to a continuous variable during training using dequantization techniques~\citep{kingma2018glow}. This is widely used by existing molecule generation methods~\citep{madhawa2019graphnvp,shi2019graphaf,liu2021graphebm}. Specifically, we add uniform noise as $\boldsymbol{\tilde{a}}_t = \boldsymbol{a}_t + \boldsymbol{u}$, $\boldsymbol{u} \sim \mathcal{U}(0,1)^p$. In the following, we elaborate how to employ flow model to construct invertible mappings $\boldsymbol{z}_t^a \rightarrow \boldsymbol{\tilde{a}}_t$, $z_t^d \rightarrow d_t$, $z_t^\theta \rightarrow \theta_t$, and $z_t^\varphi \rightarrow \varphi_t$, respectively. The training scheme is elucidated in Section~\ref{sec:training}.

% generate a
To generate $\boldsymbol{a}_t$, we first apply affine transformation to map the latent variable $\boldsymbol{z}_t^a$ to $\boldsymbol{\tilde{a}}_t$. Formally,
\begin{equation}
    \label{eq:a_flow}
    \boldsymbol{\tilde{a}}_t = \boldsymbol{\sigma}_t^a\left(\mathcal{C}^{(t-1)}\right) \odot \boldsymbol{z}_t^a + \boldsymbol{\mu}_t^a\left(\mathcal{C}^{(t-1)}\right),
\end{equation}
where the scale factor $\boldsymbol{\sigma}_t^a(\cdot) \in \mathbb{R}^p$ and the translation factor $\boldsymbol{\mu}_t^a(\cdot) \in \mathbb{R}^p$ are both dependent on the context $\mathcal{C}^{(t-1)}$. To be specific, they are computed by applying MLPs to the context-encoded representation of the selected local reference atom $f$. Formally,
\begin{equation}
\begin{cases}
    \begin{aligned}
    \boldsymbol{\sigma}_t^a\left(\mathcal{C}^{(t-1)}\right) &= \text{MLP}_\sigma^a\left(\boldsymbol{h}_f^{(t)}\right), \\
    \boldsymbol{\mu}_t^a\left(\mathcal{C}^{(t-1)}\right) &= \text{MLP}_\mu^a\left(\boldsymbol{h}_f^{(t)}\right). \\
    \end{aligned}
\end{cases}
\end{equation}
After obtaining $\boldsymbol{\tilde{a}}_t$, we can derive the one-hot $\boldsymbol{a}_t$ by performing the \textit{argmax} operation to $\boldsymbol{\tilde{a}}_t$.

% Generate d, theta, varphi
Similar to the generation of atom type, we can produce $d_t$, $\theta_t$, and $\varphi_t$ as
\begin{equation}
    \label{eq:dtv_flow}
    \begin{aligned}
    d_t &= \sigma_t^d\left(\mathcal{C}^{(t-1)}, \boldsymbol{a}_t\right) \odot z_t^d + \mu_t^d\left(\mathcal{C}^{(t-1)}, \boldsymbol{a}_t\right), \\
    \theta_t &= \sigma_t^\theta\left(\mathcal{C}^{(t-1)}, \boldsymbol{a}_t, d_t\right) \odot z_t^\theta + \mu_t^\theta\left(\mathcal{C}^{(t-1)}, \boldsymbol{a}_t, d_t\right), \\
    \varphi_t &= \sigma_t^\varphi\left(\mathcal{C}^{(t-1)}, \boldsymbol{a}_t, d_t, \theta_t\right) \odot z_t^\varphi + \mu_t^\varphi\left(\mathcal{C}^{(t-1)}, \boldsymbol{a}_t, d_t, \theta_t\right).
    \end{aligned}
\end{equation}
The scale factors $\sigma_t^d(\cdot), \sigma_t^\theta(\cdot), \sigma_t^\varphi(\cdot) \in \mathbb{R}$ and the  translation factors $\mu_t^d(\cdot), \mu_t^\theta(\cdot), \mu_t^\varphi(\cdot) \in \mathbb{R}$ are dependent on their corresponding conditional information that are defined and justified in our generation strategy, as formulated in Eq.~(\ref{eq:sequential_gen_reformulated}). These factors are also naturally parameterized by MLPs with considering their respective conditional information. To be specific,
\begin{equation}
\label{eq:mul_atom_type}
    \boldsymbol{h}_{f/c/e}^{(t)'} = \boldsymbol{h}_{f/c/e}^{(t)} \odot \text{Embedding}\left(\boldsymbol{a}_t\right),
\end{equation}
\begin{equation}
\begin{cases}
    \begin{aligned}
    \sigma_t^d\left(\mathcal{C}^{(t-1)}, \boldsymbol{a}_t\right) &= \text{MLP}_\sigma^d\left(\boldsymbol{h}_f^{(t)'}\right), \\
    \mu_t^d\left(\mathcal{C}^{(t-1)}, \boldsymbol{a}_t\right) &= \text{MLP}_\mu^d\left(\boldsymbol{h}_f^{(t)'}\right),
    \end{aligned}
    \end{cases}
\end{equation}
\begin{equation}
\label{eq:mul_dist}
    \boldsymbol{h}_{f/c/e}^{(t)''} = \boldsymbol{h}_{f/c/e}^{(t)'} \odot \text{LB}_{\text{RBF}}\left(\boldsymbol{e}_{\text{RBF}}\left(d_t\right)\right),
\end{equation}
\begin{equation}
\begin{cases}
    \begin{aligned}
    \sigma_t^\theta\left(\mathcal{C}^{(t-1)}, \boldsymbol{a}_t, d_t\right) &= \text{MLP}_\sigma^\theta\left(\left[\boldsymbol{h}_f^{(t)''}, \boldsymbol{h}_c^{(t)''}\right]\right), \\
    \mu_t^\theta\left(\mathcal{C}^{(t-1)}, \boldsymbol{a}_t, d_t\right) &= \text{MLP}_\mu^\theta\left(\left[\boldsymbol{h}_f^{(t)''}, \boldsymbol{h}_c^{(t)''}\right]\right),
    \end{aligned}
    \end{cases}
\end{equation}
\begin{equation}
\label{eq:mul_dist_angle}
    \boldsymbol{h}_{f/c/e}^{(t)'''} = \boldsymbol{h}_{f/c/e}^{(t)''} \odot \text{LB}_{\text{CBF}}\left(\boldsymbol{e}_{\text{CBF}}\left(d_t, \theta_t\right)\right),
\end{equation}
\begin{equation}
\begin{cases}
    \begin{aligned}
    \sigma_t^\varphi\left(\mathcal{C}^{(t-1)}, \boldsymbol{a}_t, d_t, \theta_t\right) &= \text{MLP}_\sigma^\varphi\left(\left[\boldsymbol{h}_f^{(t)'''}, \boldsymbol{h}_c^{(t)'''}, \boldsymbol{h}_e^{(t)'''}\right]\right), \\
    \mu_t^\varphi\left(\mathcal{C}^{(t-1)}, \boldsymbol{a}_t, d_t, \theta_t\right) &= \text{MLP}_\mu^\varphi\left(\left[\boldsymbol{h}_f^{(t)'''}, \boldsymbol{h}_c^{(t)'''}, \boldsymbol{h}_e^{(t)'''}\right]\right).
    \end{aligned}
    \end{cases}
\end{equation}
$\text{Embedding}\left(\cdot\right)$ is the same embedding layer that is used to encode ligand atom types during context encoding. Multiplying the embedding of $\boldsymbol{a}_t$ in Eq.~(\ref{eq:mul_atom_type}) helps to incorporate the generated atom type information in the subsequent generation for $d_t$, $\theta_t$, and $\varphi_t$. As the distance embedding in Eq.~(\ref{eq:schnet_variant}), $\boldsymbol{e}_{\text{RBF}}\left(d_t\right)$ is the RBF embedding of the distance $d_t$. Further, $\boldsymbol{e}_{\text{CBF}}\left(d_t, \theta_t\right)$ denotes the high-dimensional embedding of $(d_t, \theta_t)$ with circular basis functions (CBF). We use the same circular basis functions as previous works that consider geometric information~\citep{klicpera2019directional,liu2021spherical,klicpera_gemnet_2021}. $\text{LB}_{\text{RBF/CBF}}(\cdot)$ represents a linear layer and $[\cdot]$ denotes the concatenation operation. Intuitively, incorporating the distance embedding in Eq.~(\ref{eq:mul_dist}) and distance-angle embedding in Eq.~(\ref{eq:mul_dist_angle}) can guide the subsequent generation for $\theta_t$ and $\varphi_t$, respectively. This aims to capture the underlying dependencies of variables $\boldsymbol{a}_t$, $d_t$, $\theta_t$, and $\varphi_t$.

% \textbf{Overall generation process.} 

\subsubsection{Overall Generation Process}

So far, we have described the key components of our generative framework. To generate a 3D molecular geometry for a given binding site, we autoregressively place one atom at each step. At each step $t$, we firstly encode the current known context information, then select a local reference atom with our auxiliary classifiers, and finally place a new atom by producing $\boldsymbol{a}_t$, $d_t$, $\theta_t$, and $\varphi_t$ sequentially. We illustrate one generation step of our GraphBP in Figure~\ref{fig:overview_gen}. The autoregressive generation will be terminated if either no atom in the ligand can serve as a local reference atom according to the \textit{focal atom classifier} or a predefined maximum number of atoms has been achieved. Afterwards, following previous works~\citep{ragoza2021generating,luo2021sbdd}, we apply OpenBabel~\citep{o2011open} to construct bonds based on our generated 3D molecular geometries.

\subsection{Training}
\label{sec:training}
% Specifically, we have to prepare the training data, including the context $\mathcal{C}^{(t-1)}$, the local reference atoms $(f,c,e)$, the new atom type $\boldsymbol{a}_t$, and the relative location $(d_t, \theta_t, \varphi_t)$ \emph{w.r.t.} reference atoms, for each step $t$. 
To train our autoregressive generative model, we need to decompose a 3D molecule in a ligand-protein pair to a trajectory of atom placement steps. Inspired by G-SphereNet~\citep{luo2022gspherenet}, we expect that the new atom should be placed in the local region of the reference atom during generation. Thus, we select the atom in the binding site that is closest to the ligand as the first local reference atom, \emph{i.e.}, contact atom, and the atom in the ligand that is closest to the binding site as the first atom to be generated. Then, starting from this selected atom in the ligand, we apply Prim's algorithm on the 3D molecular geometry to obtain the placement order of atoms in the ligand, as well as their corresponding local reference atoms. This strategy can guarantee that the new atom for each step is always in the local region of the corresponding reference atom. With such obtained trajectory, GraphBP is trained by stochastic gradient descent using the following three loss functions.

\textbf{Atom placement loss $\mathcal{L}_{ap}$.} As described in Eq.~(\ref{eq:flow_basic}), with flow model, we can compute the log-likelihood of training data and maximize it. Hence, the training loss function for atom placement is defined as the negative of the computed log-likelihood of the training trajectory. Formally, for a 3D molecular geometry with $n$ atoms, we have
\begin{equation}
\label{eq:ap_loss}
\begin{aligned}
    \mathcal{L}_{ap} =& -\sum_{t=1}^n \left[\left(\log \left(\text{PD}\left(p_{Z_a}\left(\boldsymbol{z}_t^a\right)\right)\right)  
    + \log \left(\left |\text{PD}\left(\frac{1}{\boldsymbol{\sigma}_t^a}\right)\right |\right)\right) \right.  \\
    & + \left(\log \left(p_{Z_d}\left(z_t^d\right)\right) + \log \left(\left |\frac{1}{\sigma_t^d}\right |\right)\right)  \\
    & + \left(\log \left(p_{Z_\theta}\left(z_t^\theta\right)\right) + \log \left(\left |\frac{1}{\sigma_t^\theta}\right |\right)\right) \\
    &\left. + \left(\log \left(p_{Z_\varphi}\left(z_t^\varphi\right)\right) + \log \left(\left |\frac{1}{\sigma_t^\varphi}\right |\right)\right) \right].
    \end{aligned}
\end{equation}
$\text{PD}(\cdot)$ is used to represent the product of elements across dimensions of a vector, since $\boldsymbol{z}_t^a$ and $\boldsymbol{\sigma}_t^a$ are both $p$-dimensional vectors. Latent variables $\boldsymbol{z}_t^a$, $z_t^d$, $z_t^\theta$ and $z_t^\varphi$ can be computed by the inverted mappings of Eq.~(\ref{eq:a_flow}) and Eq.~(\ref{eq:dtv_flow}), such as $z_t^d=\frac{d_t-\mu_t^d}{\sigma_t^d}$. $p_{Z_a}$, $p_{Z_d}$, $p_{Z_\theta}$, and $p_{Z_\varphi}$ are prior Gaussian distributions. The detailed derivation of $\mathcal{L}_{ap}$ is included in Appendix~\ref{app:detailed_ap_loss}.

\textbf{Contact atom classifier loss $\mathcal{L}_{cc}$.} The \textit{contact atom classifier} is used to select the first local reference atom from the binding site. We train it with the standard binary cross entropy loss. In particular, we use the contact atom, which is the atom in the binding site that is closest to the ligand, as the positive sample, and the atom in the binding site that is furthest to the ligand, as the 
negative sample.

\textbf{Focal atom classifier loss $\mathcal{L}_{fc}$.} The \textit{focal atom classifier} is also trained with the standard binary cross entropy loss and used for selecting a local reference atom from the existing ligand atoms. The ground truth for an atom is negative if all of its bonded atoms have been generated, otherwise positive.

In summary, the overall loss function for training GraphBP is $\mathcal{L}=\mathcal{L}_{ap}+\mathcal{L}_{cc}+\mathcal{L}_{fc}$.

%   &p\left(\boldsymbol{a}_t, d_t, \theta_t, \varphi_t|\mathcal{C}^{(t-1)}\right) = p\left(\varphi_t|\mathcal{C}^{(t-1)}, \boldsymbol{a}_t, d_t, \theta_t\right) \\
%     &p\left(\theta_t|\mathcal{C}^{(t-1)},\boldsymbol{a}_t, d_t\right)p\left(d_t|\mathcal{C}^{(t-1)},\boldsymbol{a}_t\right)p\left(\boldsymbol{a}_t|\mathcal{C}^{(t-1)}\right

\section{Experiments}
\label{sec:exp}

% General intro
We firstly evaluate the ability of our GraphBP to generate 3D molecules that are capable of binding to given protein targets. The experiment demonstrates that GraphBP outperforms baselines by significant margins. Afterwards, we perform ablation studies to verify the effectiveness of the sequential generation proposed in Section~\ref{sec:step3}.

% Data intro
\textbf{Dataset.} We use the CrossDocked2020 dataset~\citep{francoeur2020three}, which contains over $22$ million docked protein-ligand crystal structures, to evaluate GraphBP for structure-based drug design. Following LiGAN~\citep{ragoza2021generating}, we ignore any poses that have root-mean-squared deviations (RMSD) greater than $2$\r{A}, thus obtaining a dataset with around $500$k protein-ligand complexes. We use the same training set and test set, as used in LiGAN, for fair comparison. Total number of atom types in ligands and in binding sites are $27$ and $19$, respectively. The atom types are summarized in Appendix~\ref{app:data_details}. 

% Post-process; metric (NOT a golden standard but reasonable)
% Justify why our evaluation metrics are convincing, given that we cannot do wet-lab experiments.

\textbf{Setup.} We use the same $10$ target proteins as LiGAN for test evaluation. Each of them could have multiple associated ligands, leading to $90$ protein-ligand pairs in the test set as reference. Following LiGAN, we generate $100$ molecules with GraphBP for each reference binding site in the test set. This evaluation setting is challenging because the test targets are diversely selected from different pocket clusters and the reference ligand usually bind strongly to the target binding site~\citep{ragoza2021generating}. We quantitatively measure the generation performance by two metrics: (i) \textbf{Validity} is the percentage of chemically valid molecules among all generated molecules. A molecule is valid if it can be sanitized by RDkit~\citep{landrum2006rdkit}. (ii) \textbf{$\Delta$Binding} measures the percentage of generated molecules that have higher \textit{predicted} binding affinity than their corresponding reference molecules. Note that we are unable to perform wet-lab experiment assays to evaluate the binding affinity of generated molecules. Also, there does not exist a computational metric that can serve as a golden standard for assessing binding affinity. Hence, following LiGAN, the binding affinity is predicted by an ensemble of CNN scoring functions~\citep{ragoza2017protein} that were trained on the CrossDocked2020 data set. Such CNN predicted affinity has been shown to be more accurate than using the Autodock Vina empirical scoring function~\citep{trott2010autodock}. Therefore, it can be used as a reasonable and convincing metric for evaluating the binding affinity of generated molecules. Following LiGAN, we firstly refine the generated 3D molecules by Universal Force Field minimization~\citep{rappe1992uff}. Afterwards, Vina minimization and CNN scoring are applied to both generated and reference molecules by using \textit{gnina}, a molecular docking program~\citep{mcnutt2021gnina}.

% Baseline (why not compare with LuoST);
\textbf{Baselines.} We consider two variants from the recent LiGAN~\citep{ragoza2021generating} method as baselines. LiGAN-prior generates molecules conditional on the given binding sites, which has the identical conditional information as our GraphBP. LiGAN-posterior encodes the whole reference protein-ligand complex as conditional information, thus generating molecules biased towards the reference molecule. Note that LiGAN-posterior incorporates more conditional information than GraphBP and LiGAN-prior. % We cannot compare with another recent work~\citep{luo2021sbdd} since its experimental setting differs from LiGAN and our GraphBP, and its implementation is not fully available as of we write this submission\footnote{Part of its implementation is available from 1/10/2022. However, the implementation is still not complete as of we submit this paper (1/27/2022).}.

\begin{table}[t]
	\caption{Generation performance on structure-based drug design. $\uparrow$ represents that higher value indicates better performance.}
	\label{tab:dock_performance}
	\centering
	\resizebox{0.8\columnwidth}{!}{
	\begin{tabular}{lcc}
		\toprule
		Method & Validity$^\uparrow$ & $\Delta$Binding$^\uparrow$  \\
		\midrule
	    LiGAN-prior &90.9\% &15.9\%\\
	    LiGAN-posterior &98.5\% &15.4\%\\
	    GraphBP (ours) &\textbf{99.7\%} &\textbf{27.0\%}\\
		\bottomrule
	\end{tabular}
	}
\end{table}

\begin{figure}[t]
	\centering
	\includegraphics[width=0.3\textwidth]{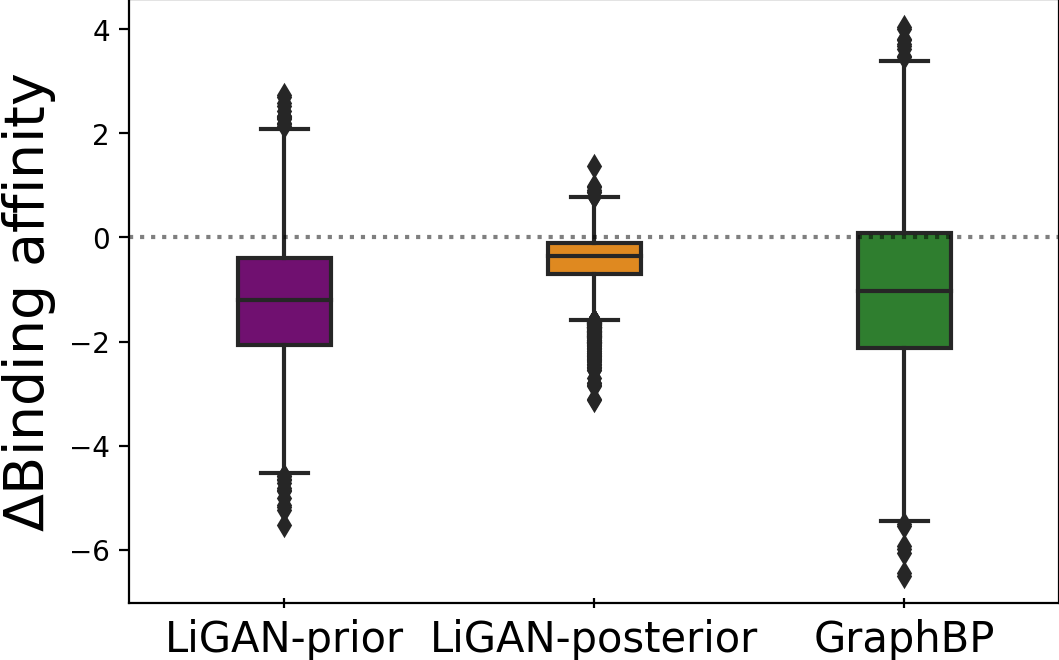}
	\caption{Visualization of $\Delta$Binding affinity distributions for LiGAN and GraphBP. The values denote the relative improvements of generated molecules over their corresponding reference molecules.}
	\label{fig:delta_binding_scores}
\end{figure}

\begin{figure*}[t]
	\centering
	\includegraphics[width=0.8\textwidth]{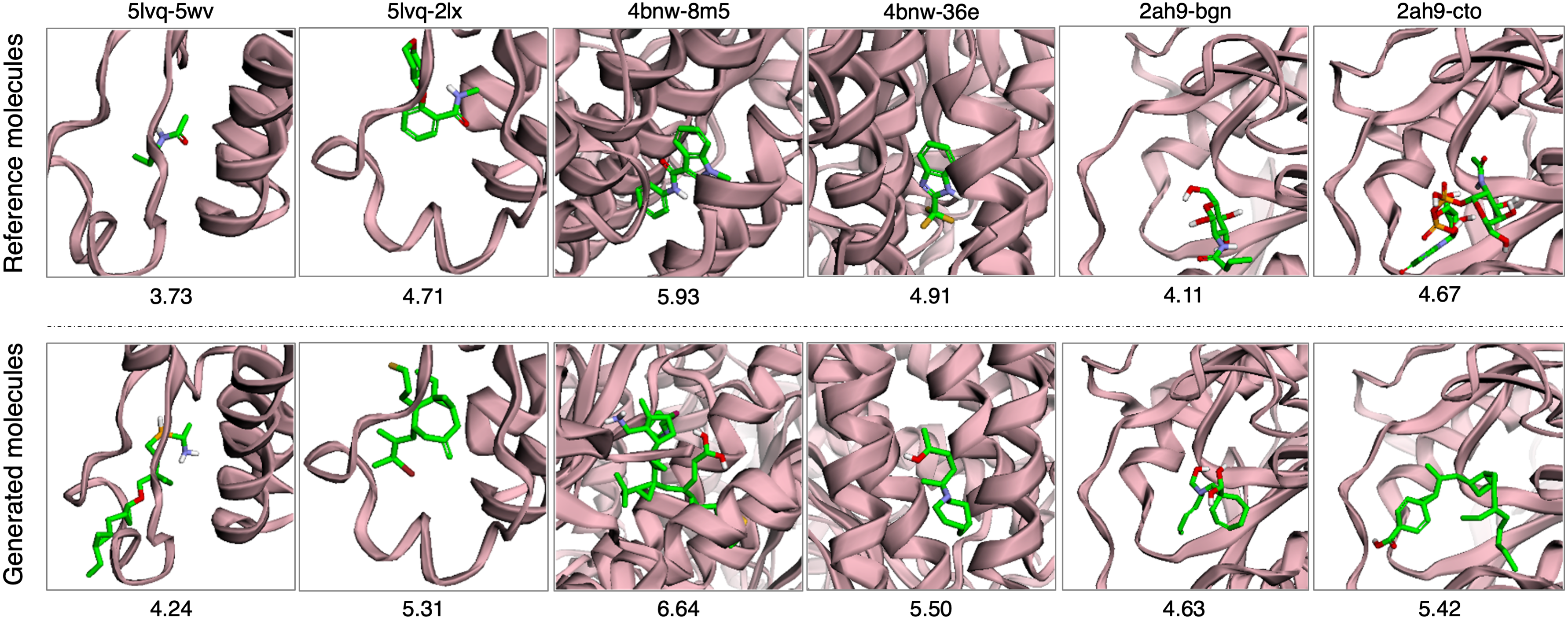}
	\caption{Several examples of generated 3D molecules that have higher predicted binding affinity than reference molecules. The PDB IDs and the ligand IDs of proteins and reference molecules are labeled on the top.}
	\label{fig:binding_visualization}
\end{figure*}

\begin{table*}[t]
  \caption{Comparison on random molecular geometry generation task between our method and ablation models. $\uparrow$ ($\downarrow$) represents that higher (lower) value indicates better performance. The top two results in terms of each metric are highlighted as \textbf{1st} and \underline{2nd}.}
  \label{tab:mmd}
  \centering
%   \resizebox{0.48\textwidth}{!}{
  \begin{tabular}{lcccccccc}
    \toprule
    \multirow{3}{*}{Method} & \multirow{3}{*}{Validity$^\uparrow$} & \multicolumn{7}{c}{MMD distances$^\downarrow$} \\
    \cmidrule{3-9}
     &  &  C-C & C-N & C-O & H-C & H-N & H-O & Avg.\\
    \midrule
    No dep. & 25.35\% & 0.776 & 0.499 & 1.251 & 2.600 & 0.823 & 2.849 & 1.466 \\
    Partial dep. & \underline{76.72\%} & \underline{0.343} & \underline{0.384} & \textbf{0.257} & \underline{0.227} &  \underline{0.373} &  \underline{0.828} & \underline{0.402} \\
    Ours & \textbf{81.98\%} & \textbf{0.232} & \textbf{0.160} & \underline{0.475} & \textbf{0.058} & \textbf{0.318} & \textbf{0.202} & \textbf{0.241} \\
    \bottomrule
  \end{tabular}
%   }
 \end{table*}

% Number + visualization
\textbf{Results.} We present the quantitative results in Table~\ref{tab:dock_performance}. Our GraphBP can generate more valid molecules than baselines, including LiGAN-posterior which even includes a valid reference ligand as conditional information. More importantly, $27.0\%$ of molecules generated by GraphBP have higher predicted binding affinity than reference molecules. This outperforms LiGAN by an absolute margin of $11.1\%$. These significant improvements over baselines demonstrate that GraphBP, which incorporates graph representations and a more flexible atom placement strategy, can capture the underlying distribution of 3D molecular geometries conditional on binding sites more effectively.

We further provide the detailed distributions of $\Delta$Binding affinity in Figure~\ref{fig:delta_binding_scores}. Note that LiGAN-posterior achieves higher average $\Delta$Binding affinity but lower variance than LiGAN-prior and GraphBP. This indicates that LiGAN-posterior, with encoding the reference molecules as conditions, might perform slight modifications on reference molecules. Even though, compared with LiGAN-posterior, our GraphBP still generates more molecules that are predicted to bind more strongly than reference molecules ($27.0\%$ \emph{vs.} $15.4\%$), demonstrating that GraphBP can generate more diverse molecules to bind with target proteins by effectively capturing the underlying conditional distribution.

In Figure~\ref{fig:binding_visualization}, we provide several examples of generated 3D molecules that are predicted to bind more strongly to the target proteins than their corresponding reference molecules. It can be observed that our generated molecules with higher predicted binding affinity are largely different from reference molecules, further indicating that our model is capable of generating diverse and novel molecules to bind target proteins, instead of simply memorizing or modifying known molecules.

% exp 2 qm9: analyze the advantage of adding dependencies, w/o dependencies --> MolGym/G-SphereNet
% (1) MMD table (2) Training loss comparison
% MMD + training loss figure + distribution comparison
\textbf{Ablation studies.} In Section~\ref{sec:step3}, we propose to generate the variables of interest sequentially to capture their underlying dependencies. Specifically, given context $\mathcal{C}^{(t-1)}$, we produce $\boldsymbol{a}_t$, $d_t$, $\theta_t$, and $\varphi_t$ one by one as $\mathcal{C}^{(t-1)} \rightarrow \boldsymbol{a}_t \rightarrow d_t \rightarrow \theta_t \rightarrow \varphi_t$. To verify the effectiveness of this strategy, we employ the following two variants. (i) \textbf{No dependencies.} The variables $\boldsymbol{a}_t$, $d_t$, $\theta_t$, and $\varphi_t$ are generated independently from the context, as $\mathcal{C}^{(t-1)} \rightarrow \boldsymbol{a}_t$, $\mathcal{C}^{(t-1)} \rightarrow d_t$, $\mathcal{C}^{(t-1)} \rightarrow \theta_t$, and $\mathcal{C}^{(t-1)} \rightarrow \varphi_t$. Thus, we omit the incorporating of atom type embedding (Eq.~(\ref{eq:mul_atom_type})), distance embedding (Eq.~(\ref{eq:mul_dist})), and distance-angle embedding (Eq.~(\ref{eq:mul_dist_angle})).  (ii) \textbf{Partial dependencies.} We consider the generated atom type information when generating $d_t$, $\theta_t$, and $\varphi_t$. However, $d_t$, $\theta_t$, and $\varphi_t$ are treated independently. It can be denoted as $\mathcal{C}^{(t-1)} \rightarrow \boldsymbol{a}_t$, $(\mathcal{C}^{(t-1)}, \boldsymbol{a}_t) \rightarrow d_t$, $(\mathcal{C}^{(t-1)}, \boldsymbol{a}_t) \rightarrow \theta_t$, and $(\mathcal{C}^{(t-1)}, \boldsymbol{a}_t) \rightarrow \varphi_t$, leading to a similar model as G-SphereNet~\citep{luo2022gspherenet}. For efficiency, we choose to conduct experiments on random molecular geometry generation, avoiding encoding large binding sites. Following G-SphereNet, we train models on 3D molecules from QM9~\citep{ramakrishnan2014quantum} and evaluate the generated molecular geometries. The evaluation metrics are validity of generated molecules and the Maximum Mean Discrepancy (MMD)~\citep{gretton2012kernel} distances of bond length distributions between generated 3D molecules and training 3D molecules. The bond length distributions of molecules generated by different models and training molecules are illustrated in Figure~\ref{fig:bond_lenth_dist}, Appendix~\ref{app:exp_details}.

\begin{figure*}[t]
	\centering
	\includegraphics[width=0.7\textwidth]{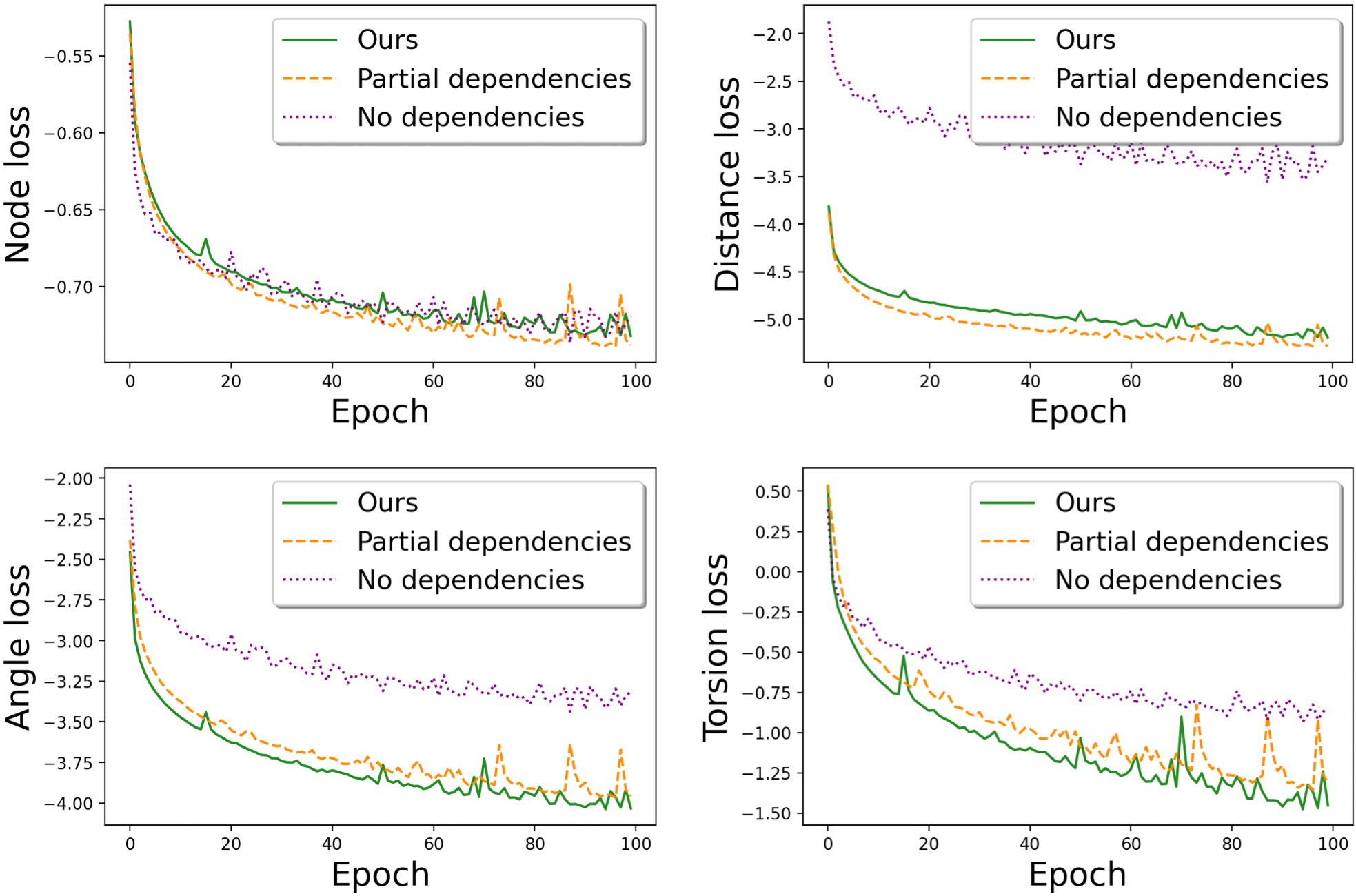}
	\caption{Comparison of training losses between our method and ablation models.}
	\label{fig:training_loss}
\end{figure*}

The comparison is summarized in Table~\ref{tab:mmd}. It shows that adding dependencies improves the generation performance consistently. Our sequential generation method performs best, demonstrating that it can model the distribution of molecular geometries more effectively by capturing the underlying dependencies among the variables. Since the loss for atom placement (Eq.~(\ref{eq:ap_loss})) can be divided into losses \emph{w.r.t.} atom type, distance, angle, and torsion, respectively, we can further analyze the modeling ability for these variables by observing their corresponding training losses. We illustrate the comparison of training losses in Figure~\ref{fig:training_loss}. By observing the loss for each variable, we can conclude that adding dependencies can help to fit the training data better.

\section{Conclusions}

In this work, we propose GraphBP, a machine learning approach to generate 3D molecules for target protein binding. GraphBP is capable of capturing 3D geometric structures and chemical interactions
of protein-ligand complexes, placing atoms without discretizing the 3D space, and preserving the equivariance property during generation. GraphBP is shown to be effective and outperforms recent baselines significantly in generating 3D molecules that bind strongly to target proteins.

\section*{Acknowledgments}

We thank David Koes and Matthew Ragoza for answering our questions about using \textit{gnina} and sharing their experimental results. This work was supported in part by National Science Foundation grants IIS-2006861 and IIS-1908220.


% \newpage 
\bibliographystyle{icml2022}
\bibliography{my_reference}

%%%%%%%%%%%%%%%%%%%%%%%%%%%%%%%%%%%%%%%%%%%%%%%%%%%%%%%%%%%%%%%%%%%%%%%%%%%%%%%
%%%%%%%%%%%%%%%%%%%%%%%%%%%%%%%%%%%%%%%%%%%%%%%%%%%%%%%%%%%%%%%%%%%%%%%%%%%%%%%
% APPENDIX
%%%%%%%%%%%%%%%%%%%%%%%%%%%%%%%%%%%%%%%%%%%%%%%%%%%%%%%%%%%%%%%%%%%%%%%%%%%%%%%
%%%%%%%%%%%%%%%%%%%%%%%%%%%%%%%%%%%%%%%%%%%%%%%%%%%%%%%%%%%%%%%%%%%%%%%%%%%%%%%
\newpage
\appendix
\onecolumn
\section{The Detailed Derivation of $\mathcal{L}_{ap}$}

\label{app:detailed_ap_loss}
The detailed derivation of $\mathcal{L}_{ap}$, as introduced in Eq.~(\ref{eq:ap_loss}), is as follows.
\begin{align}
    \mathcal{L}_{ap} &= -\log \prod_{t=1}^n p\left(\boldsymbol{a}_t, d_t, \theta_t, \varphi_t|\mathcal{C}^{(t-1)}\right) \\
    & = -\sum_{t=1}^n \log p\left(\boldsymbol{a}_t, d_t, \theta_t, \varphi_t|\mathcal{C}^{(t-1)}\right) \\
    & = -\sum_{t=1}^n \log p\left(\boldsymbol{a}_t|\mathcal{C}^{(t-1)}\right)p\left(d_t|\mathcal{C}^{(t-1)},\boldsymbol{a}_t\right)p\left(\theta_t|\mathcal{C}^{(t-1)},\boldsymbol{a}_t, d_t\right)p\left(\varphi_t|\mathcal{C}^{(t-1)}, \boldsymbol{a}_t, d_t, \theta_t\right) \\
    & = -\sum_{t=1}^n \left(\log p\left(\boldsymbol{a}_t|\mathcal{C}^{(t-1)}\right) + \log p\left(d_t|\mathcal{C}^{(t-1)},\boldsymbol{a}_t\right) + \log p\left(\theta_t|\mathcal{C}^{(t-1)},\boldsymbol{a}_t, d_t\right) + \log p\left(\varphi_t|\mathcal{C}^{(t-1)}, \boldsymbol{a}_t, d_t, \theta_t\right) \right) \label{eq:discrete_a}\\
    & \triangleq -\sum_{t=1}^n \left(\log p\left(\boldsymbol{\tilde{a}}_t|\mathcal{C}^{(t-1)}\right) + \log p\left(d_t|\mathcal{C}^{(t-1)},\boldsymbol{a}_t\right) + \log p\left(\theta_t|\mathcal{C}^{(t-1)},\boldsymbol{a}_t, d_t\right) + \log p\left(\varphi_t|\mathcal{C}^{(t-1)}, \boldsymbol{a}_t, d_t, \theta_t\right) \right) \label{eq:continuous_a}\\
    \begin{split}
    = -\sum_{t=1}^n \left[\left(\log \left(\text{PD}\left(p_{Z_a}\left(\boldsymbol{z}_t^a\right)\right)\right) + \log \left(\left|\text{PD}\left(\frac{1}{\boldsymbol{\sigma}_t^a}\right)\right|\right)\right) + \left(\log \left(p_{Z_d}\left(z_t^d\right)\right) + \log \left(\left|\frac{1}{\sigma_t^d}\right|\right)\right) \right.  \\
    \left.
    + \left(\log \left(p_{Z_\theta}\left(z_t^\theta\right)\right) + \log \left(\left|\frac{1}{\sigma_t^\theta}\right|\right)\right) + \left(\log \left(p_{Z_\varphi}\left(z_t^\varphi\right)\right) + \log \left(\left|\frac{1}{\sigma_t^\varphi}\right|\right)\right) \right].
    \end{split} \label{eq:obtained_loss}
\end{align}
$\text{PD}(\cdot)$ is used to represent the product of elements across dimensions of a vector, since $\boldsymbol{z}_t^a$ and $\boldsymbol{\sigma}_t^a$ are both $p$-dimensional vectors. Eq.~(\ref{eq:obtained_loss}) is obtained from Eq.~(\ref{eq:continuous_a}) by the property of autoregressive flow models, as described in Eq.~(\ref{eq:flow_basic}). Latent variables $\boldsymbol{z}_t^a$, $z_t^d$, $z_t^\theta$ and $z_t^\varphi$ can be computed by the inverted mappings of Eq.~(\ref{eq:a_flow}) and Eq.~(\ref{eq:dtv_flow}), such as $z_t^d=\frac{d_t-\mu_t^d}{\sigma_t^d}$. $p_{Z_a}$, $p_{Z_d}$, $p_{Z_\theta}$, and $p_{Z_\varphi}$ are prior Gaussian distributions. 

Since we apply dequantization technique to obtain $\boldsymbol{\tilde{a}}_t$ from $\boldsymbol{a}_t$ during training, the first term in Eq.~(\ref{eq:continuous_a}) maximizes the log-likelihood of $p\left(\boldsymbol{\tilde{a}}_t|\mathcal{C}^{(t-1)}\right)$ instead of $p\left(\boldsymbol{a}_t|\mathcal{C}^{(t-1)}\right)$. We have to use this dequantization technique since flow model used in our framework does not apply to discrete data directly. Note that we can simply perform \textit{argmax} operation to convert $\boldsymbol{\tilde{a}}_t$ back to $\boldsymbol{a}_t$. Hence, such dequantization can be viewed as an operation similar to data augmentation during training. Such dequantization technique is widely used and shown to be effective by existing molecule generation methods~\citep{madhawa2019graphnvp,shi2019graphaf,liu2021graphebm}.

\section{Dataset Details}
\label{app:data_details}

There are totally $27$ atom types for ligands; they are B, C, N, O, F, Mg, Al, Si, P, S, Cl, Sc, V, Fe, Cu, Zn, As, Se, Br, Y, Mo, Ru, Rh, Sb, I, W, and Au. For binding sites, there are $19$ possible atom types, including C, N, O, Na, Mg, P, S, Cl, K, Ca, Mn, Co, Cu, Zn, Se, Cd, I, Cs, and Hg.

\section{More Experimental Results}
\label{app:exp_details}

\begin{figure*}[t]
	\centering
	\includegraphics[width=\textwidth]{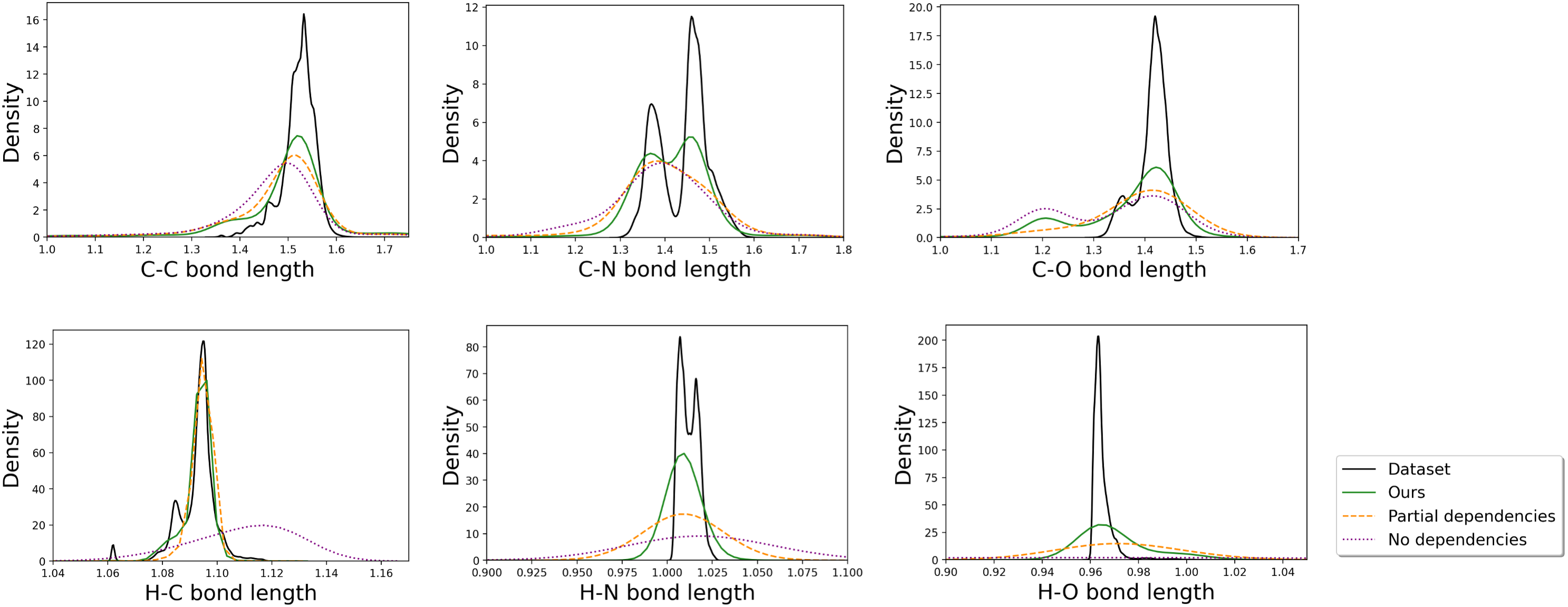}
	\caption{Visualization of bond length distributions of generated molecules and training molecules.}
	\label{fig:bond_lenth_dist}
\end{figure*}

The bond length distributions of molecules generated by different models and training molecules are compared in Figure~\ref{fig:bond_lenth_dist}. We can observe that adding dependencies among variables helps to improve the modeling ability. Our sequential generation strategy outperforms ablation variants consistently.

%%%%%%%%%%%%%%%%%%%%%%%%%%%%%%%%%%%%%%%%%%%%%%%%%%%%%%%%%%%%%%%%%%%%%%%%%%%%%%%
%%%%%%%%%%%%%%%%%%%%%%%%%%%%%%%%%%%%%%%%%%%%%%%%%%%%%%%%%%%%%%%%%%%%%%%%%%%%%%%

\end{document}